# JikesRVM: Internal Mechanisms Study and Garbage Collection with MMTk

Pradeeban Kathiravelu, Xiao Chen, Dipesh Dugar Mitthalal, Luís Veiga

{pradeeban.kathiravelu, xiao.chen, dipesh.mitthalal,luis.veiga}@tecnico.ulisboa.pt

*INESC-ID Lisboa / Instituto Superior Técnico, Universidade de Lisboa, Lisbon, Portugal.*

*Abstract* — High Level Language Virtual Machines is a core topic of interest for the researchers who are into virtual execution environments. As an open source virtual machine released to 16 universities, as early as 2001, Jikes RVM has been a major drive for many researches. While working on this project, we studied the JIT compilation of Jikes RVM as well as the Garbage Collection (GC) which is handled by the Memory Management Toolkit (MMTk), a part of the Jikes RVM. We also studied the Compressor Mark-Compact Collector algorithm and implemented it for MMTk. We have also implemented a micro-benchmark for the GC algorithms in Java, named "XPDBench", for benchmarking the implementations.

## I. INTRODUCTION

Jikes RVM is a research virtual machine for the Java language (JVM), implemented using Java. Jikes is meta-circular, as it is in the language it interprets, which is not very common. Here, the functionality of the parent interpreter is applied directly to the source code being interpreted, without any additional implementation. Hence it requires a bootstrap VM to run upon, to create a boot image. However, it doesn't run on an external JVM. Rather, a small boot image runner written using C is responsible for loading the image files at run time, and it transfers the control to the native VM code that runs on the host.

Jikes RVM is released under Eclipse Public License v 1.0 (EPL-1.0), which is an OSI approved license. It is commonly used in researches in virtual machine design, as it provides the researchers a platform to prototype the virtual machine technologies.

In the upcoming sections of this report, we will first go through the Jikes internal mechanisms we studied during this course, and then discuss the design and development of the Garbage Collection algorithm that we implemented for MMTk, followed by the evaluation and conclusion.

## II. INTERNAL MECHANISMS STUDY

Jikes RVM has extensive reading resources including research papers and presentations [1], which we used along with the code base, to understand the internal mechanisms of Jikes RVM and the related technologies.

Memory Allocation and Garbage Collection [2], and Just-In-Time Compiler were the two internal mechanisms chosen to be learned. JIT Compiler is the code generation component of a virtual machine, which compiles the byte codes into in-memory binary machine code [3], incrementally. These made us interested in learning JIT Compiler.

### 2.1 GARBAGE COLLECTION WITH MMTK

Garbage Collectors for the Jikes RVM are constructed by the framework known as the Memory Management Toolkit (MMTk) [4], which is a flexible memory management utility written using Java. MMTk outperforms GNU C's allocation algorithms by 60% on average, using different algorithms. This is due to the aggressive compiler inlining and reduced impedance because of the Java-in-Java implementation [5]. It is also ported to Rotor [6] - Microsoft's open C# runtime [7].

Key Compositional Elements in MMTk are utilities, policies, and plans. Apart from these, package *org.mmtk.harness* contains the MMTk harness test cases. Sanity checks are included in the package *org.mmtk.harness.sanity*. *GcSanity* is the collection sanity checker, that takes snapshots before and after a collection to ensure that the live objects after a collection are the same that exist before the collection. Further options are implemented in the classes of *org.mmtk.harness.options*.

### A. Mechanisms (utility)

MMTk comes with the utility classes that provides mechanism for the memory management, across multiple policies and plans that use those policies. An ideal example showing the usage of the utility package is, the interface *Constants*. All the classes either implement the Constants interface, or are sub classes of the classes that implement the interface.

In the upcoming sections, we will go through the packages, major classes and the important methods in the utility package. After going through these mechanisms, we will go through the policies and plans, accordingly.

### 1. Allocation

The package *org.mmtk.utility.alloc* handles the allocation. *Allocator* is the base class providing the basis for the processor-local allocation. This provides the retry mechanism, to prevent the slow-path allocation causing a garbage collection, violating the assumption of uninterruptibility. Allocator also ensures that requests are aligned according to requests. This class is very crucial in garbage collection as the base class for all the allocator algorithms. Improper handling of this will make it hard to trace the bugs, where allocations may cause a garbage collection, or a garbage collection may immediately follow an allocation.

The method *alignAllocation()* accepts a region to be aligned, and aligns up the request, according to the requested alignment and offset, optimizing with the known alignment for the allocator algorithms. This method returns the aligned up address.

Consecutive allocation failure for the threads are counted by *determineCollectionAttempts()*. *fillAlignmentGap()* gets the start and end addresses, and fills the region with the alignment value. The minimum size guaranteeing the allocation of a given number of bytes at the specified alignment is calculated by *getMaximumAlignedSize()*.

All allocators should use the final method, *allocSlowInline()* for the slow path allocation. This method attempts the protected method *allocSlowOnce()* multiple times, which is defined in the subclasses. This method ensures safe execution by taking care of the thread/mutator context affinity changes, whilst allowing the collection to occur.

## 2. Bump Pointer Allocation

This is implemented by the *BumpPointer* class, which extends the abstract class Allocator. Bump Pointer scans through the allocated objects linearly. To achieve parallelism, this class maintains a header at a region of 1 or more blocks granularity. The minimum region size is 32678 bytes. Hence the 3 or 4 word overhead is less than 0.05% of all space. BumpPointer is initialized by providing the space to bump point into, indicating whether the linear scanning is allowed in the region. The method *linearScan()* performs a linear scan through the objects allocated by the bump pointer, and *scanRegion()* scans through a single contiguous region.

Intra-block allocation requires only a load, addition comparison and store, and hence is fast. The allocator will request more memory, if a block boundary is encountered. The scanned objects maintain affinity with the thread which allocated the objects in the region. This class relies on the supporting virtual machine implementing the getNextObject and related operations.

Space is allocated for a new object, by calling *alloc()*. This method is frequently executed, and is sensitive to the optimizing compiler. Whenever the bump pointer reaches the internal limit, *allocSlow()* is called. This method should never be inlined by the optimizing compiler, and hence is annotated with *@NoInline,* to force out of line.

Bump pointer can be re-associated to a different space by calling *rebind()*, providing the new space to which the pointer to be associated to. The bump pointer will be reset such that it will use the new space for the next call to *alloc()*.

*Address* is a stub implementation of an Address type, used by the runtime system and collector to denote the machine addresses. An allocation unit is denoted as a *card,* which is marked by an address that lies within the card. Providing the address of the object creating a new card, the address that lies within the card, and the size of the pending allocation in bytes, *createCardAnchor()* creates a record, where the start of the card is relative to the start of the object. The start of the card corresponding to the given address can be retrieved by calling *getCard()*. Similarly, the address of the card metadata can be retrieved by providing an address from the card, calling *getCardMetaData()*.

Next region from the linked list of regions can be retrieved using *getNextRegion()*. Similarly, *setNextRegion()* is used to set the next region in the linked list. *clearNextRegion()* clears the next region pointer in the linked list.

The DATA_END address from the region header can be retrieved using *getDataEnd()*, by providing the bump pointer region address. *setDataEnd()* is used to set the new DATA_END address from the header. Similarly, *getRegionLimit()* and *setRegionLimit()* return or store the end address of the given region, respectively. The lowest address where the data can be stored can be retrieved by getDataStart(), for the given region. *updateMetaData()* is used to update the metadata, reflecting the addition of a new region. Where a bump pointer region has been consumed, but the contiguous region is available, *consumeNextRegion()* consumes it and returns the start of the region satisfying the outstanding allocation request.

## 3. Block Allocation

Blocks are a unit of storage of $2 \wedge n$ bytes, that are non-shared (thread-local) and coarse-grained. Virtual memory resource provides virtual memory space. Here, pages consumed by blocks are accountable for the memory resource. *BlockAllocator* implements various sizes of Block data structure. *alloc()* allocates a block and returns the first usable bytes of the block. A block is freed by calling *free()*. If the block is completely free, the block is returned to the virtual memory resource. Otherwise, if the block is just a sub-page block, the block is added to the free list.

## 4. GCspy Integration

GCspy 2.0 [8] [9] is a tool that helps to analyse the heap, that often is used to understand the memory usage and effectiveness of garbage collectors in our project. The development of GCspy however lags behind that of the Jikes RVM core [10]. The data for GCspy is gathered using *gcspyGatherData()* in the classes. The package *org.mmtk.utility.gcspy* contains the classes for the GCspy integration, and *org.mmtk.utility.gcspy.drivers* contains the GCspy drivers for the MMTk collectors.

## 5. Treadmill

Each instance of the *Treadmill* is a doubly linked list, where each node is a piece of memory. The words of each node,

```
[Forward Link | Backward Link | Treadmill | Payload ----->]
```

The Treadmill object must not be moved. Depending on the constructor argument, access to the instances may be synchronized.

This assumes that the implementation language, and the language being implemented are the same. This works well with Jikes RVM, as these are Java. However, for using with other VM implementations in other languages, the doubly linked list class, upon which Treadmill depends, should be rewritten.

If the created instance is to be shared between the threads, which is decided by the boolean parameter "*shared"* provided as a parameter in the constructor, the access will be synchronized with locks. A node is added to the treadmill during the allocation, using *addToTreadmill()*.

### 6. Large Object Allocation

This is implemented in *LargeObjectLocal,* which extends the abstract class *LargeObjectAllocator.* Each instance provides a fast unsynchronized access to the treadmill, and bound to a single CPU. Hence this should not be shared across different CPUs, as they provide truly concurrent threads.

Given c CPUs, and t Treadmill spaces, there will be c * t instances of this class, for each {CPU, Treadmill} pair.

### 7. Page Resource

Allocation of pages for a space is managed by the abstract class *PageResource*. When a space request a page, page budget and the use of virtual address space are checked. If the request can not be satisfied, garbage collection is triggered. *MonotonePageResource*, which is a subclass of this class handles the monotonic space usage. Similarly, the other sub class, *FreeListPageResource* handles the ad hoc usage. Copying collectors allocate space monotonically before freeing the entire space. Hence, MonotonePageResource is useful for them. Though the MonotonePageResource is more restrictive, it is easier to manage.

### 8. Heap Growth Management

HeapGrowthManager observes the heap utilization and GC load, and grows and shrinks the heap size accordingly. This class, and all the other classes in the package *org.mmtk.utility.heap* contains the heap related mechanisms.

### 9. Sanity Checker

Sanity checks for the simple collectors are handled by the classes in the package *org.mmtk.utility.sanitychecker*. *SanityChecker* is the major class handling the sanity checking. *SanityDataTable* implements a simple hashtable to store and retrieve per object information on sanity checks.

### 10. Statistics

The package *org.mmtk.utility.statistics* contains a number of counters, implementing the abstract class *Counter* for multiple counting purposes. *SizeCounter,* aggregates two *EventCounter* objects*,* to count the number of events and the volume. *Xml* class writes the output in XML format.

### 11. Deque

The package *org.mmtk.utility.deque* defines the doubly-linked, double-ended queue (deque). Though the double linking slightly increases the space demand, this is a worth trade off, as this provides an efficient buffer and operations such as sorting.

*LocalSSB* implements a local unsynchronized sequential store buffer. This is used in critical code such as garbage collection work queue and the write buffer used by many collectors. Hence, this is implemented as efficient as possible in space and time. Each instance has a bump pointer and a pointer to the sharedDeque. This class follows FIFO, though it doesn't implement dequeing. *TraceBuffer* supports enqueuing and dequeuing of tracing data and bulk processing of the buffer.

### 12. Log

*Log* class is used for trace and error logging. Message buffer size is intentionally kept large (3000 characters), as *Lock* class of Jikes RVM logs a considerable amount of information during a potential GC deadlock.

## B. Policies (policy)

Spaces are contiguous regions of virtual memory, that is managed by a single policy. MMTk maps policies to spaces. In an address space, any given policy can manage multiple spaces. Policies are implemented following the local/global pattern, and named XXXSpace and XXXLocal. Here, XXXSpace refers to any of the classes that extend the abstract class, Space, or its sub classes. XXXLocal refers to any classes that extend the abstract class Allocator, or its sub classes.

1. Copying Collector – (*CopySpace | CopyLocal*)
2. Explicitly Managed Collector – (*ExplicitFreeListSpace | ExplicitFreeListLocal*)
3. Immortal Collector – (*ImmortalSpace | ImmortalLocal*)
This doesn't actually collect, nor does it hold any state. Collector just propagates marks in a liveness trace.
4. Treadmill Collector – (*LargeObjectSpace | LargeObjectLocal*)
Each instance corresponds to one explicitly managed large object space.
5. Mark-Compact Collector (*MarkCompactCollector*) – (*MarkCompactSpace | MarkCompactLocal*)
6. Mark-Sweep Collector – (*MarkSweepSpace | MarkSweepLocal*)

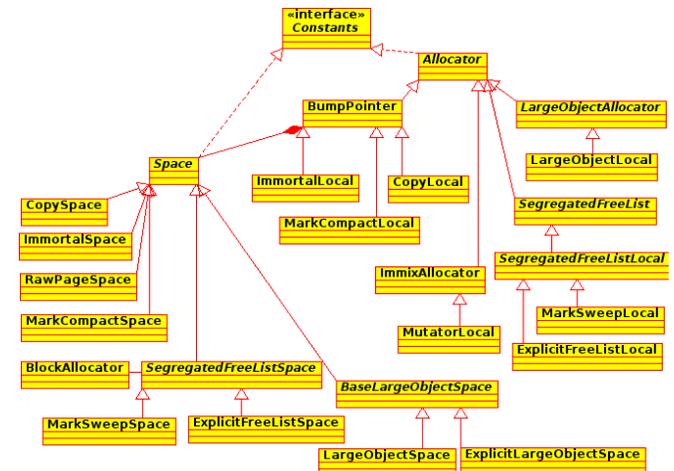

Fig. 1 Class Diagram of Policies

Each instance of a policy space maps to a single virtual memory space, with P instances of the local class attached, where the collector is P-way parallel. As a global object, space is shared among multiple mutator threads, where the XXXLocal instances are created as part of a mutator context inside the respective XXXMutator class.

This Space|Local pattern lets the local Allocator allocate space from a thread-local buffer till it is exhausted, and after that letting it allocate a new buffer from the global Space, with appropriate locking. The constructor of the XXXLocal

specifies the space that the allocator will allocate global memory from.

## C. Plans (plan)

*Plan* is the highest level of composition, as it composes policies to build a memory management algorithm. Collector algorithms are implemented in policies and plans. Referring to the respective collector algorithms from the papers is the ideal way to understand each of these polices/plans.

*build/configs* folder contains multiple properties files, which can be used to build Jikes RVM with different configurations and different collectors. MMTk plan is defined in the properties files. The recommended production or prototype configurations use GenImmix plan, as can be seen from *production.properties* and *prototype.properties*.

```
config.mmtk.plan=org.mmtk.plan.generational.im
mix.GenImmix
```

### 1. Plan

*Plan* is the abstract class that defines the core functionalities of all the memory management schemes, and hence becomes the base class for all the collector plans, as shown in Fig. 2.

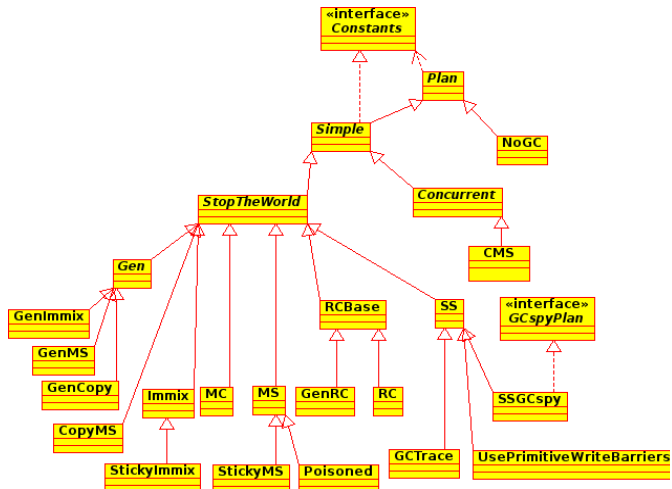

Fig. 2 Class Diagram of Plans

Global and local states are separated into different classes, as in the case of policies, as plans make a clear distinction between the global and thread-local activities. Global activities should be synchronized, where this is not required for thread-local activities. Static resources such as VM and memory resources are defined and managed by the global instances.

*ParallelCollectorGroup* is a pool of collector contexts, that can be triggered to perform collection. Plan consists two instances of this, named parallelWorkers and concurrentWorkers.

The respective sub class of Plan will have a single instance and there will be a one-to-one mapping between the PlanLocal and CPUs (kernel threads). This mapping is crucial in understanding the properties of the plans. Due to this separation, instance methods of the Local let the functions such as allocation and collection unsynchronized and faster.

### 2. Boot Sequence

The boot sequence calls the methods given below in the given order.

1. *enableAllocation()* - called early in the boot process for the allocation.
2. *processOptions()* - called immediately by the run time, as the command line arguments are available. Calling the method *enableAllocation()* is a precondition, as the infrastructure may require allocation to be able to parse the arguments. All plans operate in the default minimum, till this method is called, and the respective values are obtained.
3. *enableCollection()* - called when it is safe to spawn contexts and start garbage collection. If the threads have not explicitly been set, right defaults will be used. Creates the parallel workers and concurrent workers. Creates the control thread and allows mutators to trigger the collection.
4. *fullyBooted()* - called just before giving the control back to the application, and hence marks the completion of the collection cycle.

Before the VM exists, *notifyExit()* is called for the clean up. Further, plan specific timing information can be provided by overriding *printDetailedTiming()*. *collectionPhase()* performs a global collection phase.

### 3. Nursery

*Nursery* refers to the collection of objects allocated since last GC, hence a collection of short living objects. Generational collectors function based on the observation that an object living longer than a specific amount of time will live further long. That means, when a few objects in the nursery survives a few round of collections in the nursery, they are moved to the matured space. Matured space consists of 90% of space, but of 10% of the number of objects that are created from the beginning. Nursery contains just 10% of the heap space, but contains 90% of the objects created ever since, as the objects in nursery are often short-lived. Matured space is scanned rarer than the nursery in the generational collector algorithms, where the matured space is scanned only when the specific time occurs, or when the VM runs out of memory. *IsCurrentGCNursery()* checks whether the current GC is only collecting the objects allocated after the last garbage collection.

After a full heap GC, *sanityLinearScan()* performs a linear scan of all spaces for possible leaks. Triggering of a garbage collection is controlled by *collectionRequired()*. This is called periodically during the allocation. If collection is requested by the plan, this method would return true. Whether the object can ever move is determined by the method *willNeverMove()*.

### 4. MutatorContext

*MutatorContext* and its subclasses define the per-mutator thread behaviour. MutatorContext is the base class of all the per-mutator contexts, ConcurrentMutator, StopTheWorldMutator, and all the XXXMutator (GC-specific sub classes), where XXX refers to the names of the collector plans such as GenImmix or CMS.

This implements the basic unsynchronized per-mutator behaviour that is common to all the collectors such as the support for immortal and large object space allocation and empty stubs for write barriers.

*initMutator()* is called before the MutatorContext is used. It is called after the context is fully registered and visible to the collection. It notifies that the context is registered such that it will be included in the iterations over the mutators. Similarly, *deinitMutator()* is the final method to be called, where the mutator is to be cleaned, and hence all local data should be returned. The implementation of deinitMutator() in this class, as well as the sub classes of MutatorContext by default, call the *flush()*, which flushes the mutator context, flushing the remembered sets into the global remset pool.

*checkAllocator()* performs a run time check for the allocator.

## 5. CollectorContext

*CollectorContext* is an abstract class that is extended by the classes XXXCollector, similar to the Mutators. These classes implement the per-collector thread behaviour and structures such as collection work queues.

MMTk assumes that for each thread participating in the collection, the VM instantiates instances of CollectorContext and MutatorContext in thread local storage (TLS). Hence the access to this state is assumed to be low-cost at GC time (opposed to the access to MutatorContext, which is low-cost during mutator time), essentially separating this class which is thread-local, from the global operations in the Plan class. Operations in this class and the subclasses are local to each collector thread, where operations in the MutatorContext class and all of its sub classes are local to each mutator thread, whilst the synchronized operations are via the access to Plan and its subclasses, which are global. Synchronization is localized, explicit, and thus minimized.

Memory is allocated for an object using *alloc()*. Any required post allocation operations are handled by *postAlloc()*. *getAllocatorFromSpace()* finds the allocator that is associated with the given space.

The number of active collector threads (n), which is often the number of processors and the number of mutator (application) threads (m) are decided by the VM, and not MMTk. Collector operations are separated as below.

Per-collector thread operations: The operations for the entire GC, performed by each collector thread, and m per-mutator thread operations multiplexed across the n active collector threads.

Per-mutator thread operations: The operations such as flushing and restoring per-mutator state.

The unpreemptible method *run()* provides an entry point for the collector context.

Space is allocated for copying the space using *allocCopy()*. This method just allocates the space, and doesn't really copy the object. Actions after the copy are performed by *postCopy()*. Run time check of the allocator for the copy allocation is performed by *copyCheckAllocator()*.

We often need to find the number of parallel workers currently executing to be able to load balance better. This can be found by calling *parallelWorkerCount()*, from anywhere within the collector context.

## 6. NoGC

This simple No-garbage-collector class implements the global state of a simple allocator without a collector.

## 7. Simple

The abstract class Simple provides the core functionalities of a Simple collector, which should be extended by the two types of collectors, non-concurrent or concurrent. Collection phases are defined along with their base level implementations. The spaces introduced by the subclasses should be implemented by them respectively.

## 8. StopTheWorld

All the non-concurrent collectors should extend the abstract class *StopTheWorld*.

## 9. Concurrent

The abstract class *Concurrent* implements the global state of a concurrent collector.

## 10. Collector Plans

1. CMS (concurrent.marksweep)
2. CopyMS (copyms)
3. GenCopy (generational.copying)
4. GenMS (generational.marksweep)
5. GenImmix (generational.immix)
6. MC (markcompact)
7. MS (marksweep)
8. RC (refcount.fullheap)
9. GenRC (refcount.generational)
10. SS (semispace)
11. GCTrace (semispace.gctrace)
12. UsePrimitiveWriteBarriers (semispace.usePrimitiveWriteBarriers)
13. StickyImmix (stickyimmix)
14. StickyMS (stickyms)

### 2.2 JUST-IN-TIME (JIT) COMPILER

Jikes RVM includes two compilers. One is the baseline compiler, and the other is optimizing compiler, which is commonly referred to as the JIT compiler. Here, we will study the optimizing compiler in Jikes RVM. The package *org.jikesrvm.compilers* contains the source code of the compilers, where The package *org.jikesrvm.compilers.opt* contains the code of the optimizing compiler.

*SharedBooleanOptions.dat* found in the directory */rvm/src-generated/options* defines the boolean options for the optimizing compiler [11]. Similarly, *SharedValueOptions.dat* in the same directory defines the non-boolean options.

## 1. Transformations of Methods

Jikes RVM takes methods as the fundamental unit of optimization, and optimize them by transforming the intermediate representation as below.

```
Byte Code → [Optimizing Compiler] → Machine
Code + Mapping Information
```

Mapping information consists of garbage collection maps, source code maps, and execution tables [12]. Optimizing compiler has the following intermediate phase transitions, for the intermediate representation.

```
High-level Intermediate Representation (HIR)
→ Low-level Intermediate Representation →
Machine Intermediate Representation.
```

The general structure of the master plan consists of elements,
1. Converting, *byte codes* → *HIR*, *HIR* → *LIR*, *LIR* → *MIR*, and *MIR* → *Machine Code*.
2. Performing optimization transformations on the HIR, LIR, and MIR.
3. Performing optimization.

## 2. CompilationPlan

*CompilationPlan* is the major class, which instructs the optimizing compiler how to optimize a given method. Its constructor constructs a compilation plan, as the name indicates. This includes the instance of *NormalMethod,* the method to be optimized. An array of *TypeReference* is used in place of those defined in the method.

This also contains an object of *InstrumentationPlan,* defining how to instrument a method to gather the run time measurement information. The methods *initInstrumentation()* and *finalizeInstrumentation()* are called at the beginning of the compilation, and after the compilation just before the method is executed, respectively.

A compilation plan is executed by executing each element in the optimization plan, by calling the method *execute()*. The compiler queries the *InlineOracle* interface to decide whether to inline a call site. If there is some instrumentation to be performed, initialization takes place. After the instrumentation, finalization takes place to clean up. However, the finalization will not be executed, if this fails with an exception.

## 3. OptimizationPlanElement

An array of *OptimizationPlanElement* defines the compilation steps. OptimizationPlanElement thus represents an element in the optimization plan. Instances of the sub classes of the abstract class OptimizationPlanElement are held in *OptimizationPlanner.masterPlan*, and hence they represent the global state. It is incorrect to store any per-compilation state in the instance field of these objects. OptimizationPlanner specifies the order of execution during the HIR and LIR phases for a method. The method *reportStats()* generates a report on time spent performing the element.

The method *shouldPerform()* determines whether the optimization plan should be performed and be part of the compilation plan, by consulting the passed OptOptions object. When an element is included, all the aggregate elements that the element is a component of, are included too. The work represented by the element is done in the optimization plan, by *perform()*. The work done is assumed to modify the Intermediate Representation (*IR*) in some way. The perform() of the aggregate element will invoke all the elements' perform().

A client is a compiler driver that constructs a specific optimization plan by including all the *OptimizationPlanElement*s from the master plan, that are appropriate for this compilation instance.

## 4. OptimizationPlanAtomicElement

The final class *OptimizationPlanAtomicElement* extends OptimizationPlanElement. This object consists of a single compiler phase in the compiler plan. The main work of the class is done by its phase, which is an instance of the class *CompilerPhase*. Each phase overrides this abstract class, and specifically the abstract methods *perform(),* which does the actual job, and *getName(),* which gets the name of the phase. A new instance of the phase is created when shouldPerform() is called, and is discarded as soon as the method is finished. Hence, the per-compilation state is contained in the instances of the sub classes of CompilerPhase, as the state is not stored in the element.

## 5. OptimizationPlanCompositeElement

Similarly, *OptimizationPlanCompositeElement* is the base class of all the elements in the compiler plan that aggregates together the other OptimizationPlan elements, as depicted in Fig. 3.

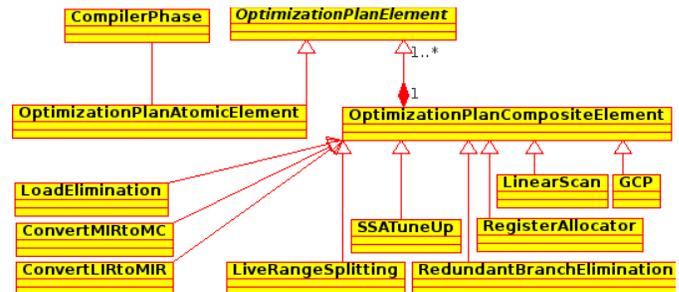

Fig. 3 Class Diagram of OptimizationPlanElement

Here, an array of OptimizationPlanElement is kept to store the elements that compose this composite element. The constructor composes together the elements passed through as a method parameter, into the composite element. If the phase wants the IR to be dumped before or after it runs, it can be enabled using *printingEnabled().* By default, this is disabled, where the sub classes are free to override this method, and make it true.

## 6. Sub classes of OptimizationPlanCompositeElement

**6.1. LinearScan** – Handles the linear scan register allocation.

**6.2. GCP** – Handles Global Code Placement. Comes in two variants. One is, *Loop Invariant Code Motion (LICM)*, which is applied to HIR and LIR. The other is, *Global Common Sub Expression Elimination (GCSE)*, which is applied only to LIR and before LICM. These run on static single assignment form (SSA form). SSA form is a property of IR that states that each variable is assigned exactly once. Utility functions of SSA form are handled by the class *SSA*. These GCP algorithms use the dominator tree to determine the positions for operations. However, SSA, by default is disabled in the optimization compiler, as it is currently buggy.

**6.3. SSATuneup** – Places the IR in SSA form, and cleans up by performing a set of simple optimizations.

**6.4. RegisterAllocator** – The driver routine for register allocation. The constructor initially prepares for the allocation, and then performs the allocation using the live information.

**6.5. LoadElimination** – Implements the redundant load elimination.

**6.6. LiveRangeSplitting** – Performed at the end of SSA in LIR. This performs live range splitting, when they enter and exit the loop bodies by normal/unexceptional control flow, also splitting on edges to and from infrequent code.

**6.7. RedundantBranchElimination** – Based on the SSA form, global value numbers, and dominance relationships, the below conditions are considered to be sufficient for a branch to be eliminated as redundant.

* It has the same value number as another conditional branch, and hence it is found to be equivalent to that branch.
* Target of taken branch of the other condition branch (cb2) dominates the condition branch (cb1) under consideration, making the target block having exactly one in edge.
* Non-taken continuation of cb2 dominates cb1, and the continuation block has exactly one in edge.

**6.8. ConvertLIRtoMIR** – Converts an IR object: LIR → MIR, using the Bottom-Up Rewrite System (*BURS*), which is a tree pattern matching system, for instruction selection. The constructor proceeds to create the phase element as a composite of the other elements, in stages as given below.

* Reduce the LIR operator set to a core set.
* Convert ALU operators.
* Normalize the usage of constants.
* Compute the liveness information, build DepGraph block by block and perform BURS based instruction selection.
* Complex operators expand to multiple basic blocks of MIR. Handle them.
* Perform Null Check Combining using validation operands. Remove all the validation operands that are not present in the MIR.

**6.9. ConvertMIRtoMC** – Converts an IR object: MIR → Final Machine Code. In the constructor, it initially performs the final MIR expansion. Then, it performs the assembly and map generation.

## 7. Subclasses of CompilerPhases

Different optimizations of different phases are taken care by the sub classes of CompilerPhases, which are responsible for them. Different optimization plans of different functions will still share the same component element objects, as every optimization plan contains a sub set of the elements from the master plan. To avoid this overhead, *perform()* of the atomic element creates a new instance of the phase immediately before calling the phase's *perform()*. The method *newExecution()* of *CompilerPhase* is overridden, such that, if the phase doesn't contain a state, the phase is returned instead of its clone.

## 8. Operators

IR includes a list of instructions, each including an operand or possibly, the operands. Operators are defined by the class *Operator,* which is auto-generated by the driver, from a template. The input files are *Operators.Template* and *OperatorList.dat* found in *rvm/src-generated/opt-ir*. This machine dependent *Operator* class resides in *jikesrvm/generated/ {ARCH}/main/java/org/jikesrvm/compilers/opt/ir,* where ARCH refers to the architecture such as ia32-32, ia32-64, ppc-32, and ppc-64.

*OperatorList.dat* in *rvm/src-generated/opt-ir* defines the HIR and LIR subsets of the Operators where the first few define HIR only whilst the remaining define both, and *OperatorList.dat* in *rvm/src-generated/opt-ir/{ARCH}* (Here, ARCH is such as ia32) defines the MIR subset. Each operator definition consists of 5 lines given below:

*SYMBOL* – A static symbol identifying the operator.

*INSTRUCTION_FORMAT* – The class of Instruction format that accepts this operator.

*TRAITS* – Characteristics of the operator, composed with a bit-wise OR | operator. Valid traits are defined in the Operators class.

*IMPLDEFS* – Registers implicitly defined by the operator.

*IMPLUSES* - Registers implicitly used by the operator. Here the last two are usually applicable only to the MIR subset, which is machine dependent.

For example, the definition of the Floating Point Conditional Move (FCMOV) is,

```
IA32_FCMOV
MIR_CondMove
none
C0_C1_C2_C3
CF_PF_ZF
```

where, for the integer addition operator, it is,

```
INT_ADD
Binary
commutative
```

leaving the last two lines empty.

## 9. Instruction Formats

Instruction formats are defined in the package instructionFormats and each fixed length instruction format is defined in the *InstructionFormatList.dat* files in */rvm/src-generated/opt-ir* and */rvm/src-generated/opt-ir/{ARCH}* (Here, ARCH is such as ia32 and ppc).

Each entry in the InstructionFormatList.dat has 4 lines as below.

*NAME* – Name of the Instruction Format.

*SIZES* – Parameters such as the number of operands defined (*NUMDEFS*), defined and used (*NUMDEFUSES*), and used (*NUMUSES*), and additionally *NUMVAR, VARDORU,* and *NUMALT* for variable length instructions.

*SIG* – Describing the operands. The structure is, *D/DU/U NAME TYPE [opt]*. Here, D/DU/U defines whether the operand is a def, def and use (both), or use. *Type* defines the type of the operand, as a sub class of *Operand*. *[opt]* indicates whether the operand is optional.

*VARSIG* – Describing the repeating operands, used for variable length instructions. The structure is *NAME TYPE [PLURAL]*. Here, PLURAL is used, where *NAME*s is not the plural of *NAME*.

For example, let's consider the definition of NEWARRAY.

```
NewArray
1 0 2
"D Result RegisterOperand" "U Type TypeOperand" "U Size Operand"
```

Here it indicates the three operands, one D (def), no operand of DU (def and use) type, and two U (use), where the Def type operand is called *Result,* and of type RegisterOperand, similarly the use operands are called *Type* and *Size*, and are of types *TypeOperand* and *SizeOperand,* respectively.

**10. Instruction Selection**

BURS is used for instruction selection, where the rules are defined in architecture specific files in *rvm/src-generated/opt-burs/{ARCH}*, where *ARCH* refers to architectures such as ia32 (example files: PPC_Alu32.rules, PPC_Alu64.rules, PPC_Common.rules, PPC_Mem32.rules, and PPC_Mem64.rules) or ppc (example files: IA32.rules, IA32_SSE2.rules, and IA32_x87.rules).

Each rule is defined by an entry of four lines, as given below.

*PRODUCTION* – The format is "*non-terminal: rule*", where this is the tree pattern to be matched. *Non-terminal* denotes a value, followed by a colon, and followed by a dependence tree producing that value.

*COST* – The additional cost due to matching the given pattern.

*FLAGS* – Defined in the auto-generated class *BURS_TreeNode*. The below can be the flags.

 * *NOFLAGS* – No operation performed in this production.
 * *EMIT_INSTRUCTION* – Instructions are emitted in this production.
 * *LEFT_CHILD_FIRST* – The left hand side of the production is visited first.
 * *RIGHT_CHILD_FIRST* – The right hand side first.
 * *TEMPLATE:* The code to be emitted.

Floating point load is defined below as an example.

```
Fpload: FLOAT_LOAD(riv, riv)
0
EMIT_INSTRUCTION
pushMO(MO_L(P(p), DW));
```

Jburg [13], a bottom-up rewrite machine generator for Java, is a compiler construction tool, that is used for instruction selection, converting the tree structure representation of the program into a machine code.

### III. DESIGN AND IMPLEMENTATION

While building the platform, we were able to start working on a few bugs in the software, and had already issued a patch to fix a minor bug [14], which was also committed to the trunk. Similarly, we continued learning the internal mechanisms, even during the latter phases, whenever needed.

Implementation of the Compressor Mark-Compact Algorithm [15] to MMTk's suite of collectors for the garbage collection, is a project proposed by the community. This concurrent, incremental, and parallel compaction algorithm, known as the "Compressor", uses a single heap-pass, while preserving the orders of the object, utilizing the entire heap space, while solving the fragmentation issues, which makes it an attractive algorithm for research and implementation. An initial implementation effort [16] was done along this line, but was not completed. We will now analyse the development and testing plans further, in this section.

#### A. TIMELINE - INCREMENTAL DEVELOPMENT STRATEGY

The algorithm is,

1. Parallel – On multiple processors, compaction can be done in parallel. Mutator threads are suspended, when the memory is exhausted. Collector threads are activated and perform collection in parallel. Upon the completion, the collector threads are suspended, and the mutator threads resume.
2. Incremental – Most of the parallel compaction can be run incrementally, by the application threads. Some compact work is done during each allocation.
3. Concurrent – Whilst the application is executing, part of the collection can run concurrently. Here the collector threads can be run along with the mutator threads exclusively, or in addition to the stop-the-world phases that we implement in the first phase of implementation.

We saw these properties as an opportunity for an incremental development, deployment and testing, which we find promising and suitable as our group project. An incremental version of the algorithm, which is not concurrent, was developed initially and tested to ensure that it functions as a parallel compactor. The development of the concurrent version was started at the latter part of the development cycle.

#### B. ALGORITHM – DESIGN AND IMPLEMENTATION

Now we will look into the design and implementation of the Compressor, by going through each of the functionalities.

**1. Compressor**

*Compressor*, the major class of implementation is based on *SS* (SemiSpace) implementation. This implements the global state of the Compressor collector. It is a singleton class that is a subclass of the abstract class *Plan*. As the class that encapsulates the data structures shared among multiple threads, three doubly linked lists are defined in the Compressor as depicted in Fig. 4, that serve as the means of work sharing, between *CompressorMutators* and *CompressorCollectors* and between different stages of *CompressorCollector*.

```
Compressor
- isSwapped : boolean = false
- ready : boolean = false
- compressorCollection : short
- calculateHeads : DoublyLinkedList
- preallocCopyHeads : DoublyLinkedList
- allocationHeads : DoublyLinkedList
+ collectionPhase(phaseId : short)
+ willNeverMove(object : ObjectReference)
# registerSpecializedMethods()
+ getPagesRequired() : int
+ getPagesAvail() : int
+ getPagesUsed() : int
```

Fig. 4 Compressor Class

```
/* Mutators push the initial addresses of a
BumpPointer and Collectors pop these
addresses and use them to traverse the
regions.*/
public final DoublyLinkedList calculateHeads
= new DoublyLinkedList(LOG_BYTES_IN_PAGE,
true);

/* Sharing of region addresses between the
```

```
stages, CALC_METADATA and PREALLOC_COPY.*/
public final DoublyLinkedList
preallocCopyHeads = new
DoublyLinkedList(LOG_BYTES_IN_PAGE, true);

/* Sharing of region addresses between the
collector and the mutator that runs after a
collection.*/
public final DoublyLinkedList allocationHeads
= new DoublyLinkedList(LOG_BYTES_IN_PAGE,
true);
```

### 2. Virtual Spaces

Compressor holds two instances of *MarkCopySpace*. These two instances are considered virtual address spaces, as they are not always mapped to the physical addresses. These are of the same size as heap, and also switch their role as *to-virtual-space* and *from-virtual-space*, during the execution. During each compaction objects are moved from the from-virtual-space to to-virtual-space, and the roles are switched after that. This switching of the roles and the current role of each MarkCopySpace is tracked by the boolean entry *isSwapped*, which is initially set to false, and keeps changing from false to true and vice versa, when the roles are switched. Compressor also holds a Deque. This queue is used for sharing the BumpPointer region anchors. The collector will take them from this queue.

### 3. Stages

Compressor algorithm has three major stages. In the *Mark stage* it marks the objects in a side markbit. In the *compute stage*, it computes the offset and first object vectors. In the *final stage*, the compaction threads copy the objects to compact.

### 4. Compressor Utilities

The package *org.mmtk.policy.compressor.util* contains the classes that contain the utility methods of Compressor policy. *MetadataUtil* holds the utility methods needed to access chunk metadata. *Chunk* refers to Space chunks (usually 4MB).

The general memory map assumes the structure given below, where Region Meta data (BumpPointer Metadata) consists of Region End, Next Region, and Data End.

```
[Region End| Next Region| Data End| Data -->]
```

In the methods such as *testAndSetLiveBits()* or *clearLiveBits()* in MetadataUtil, only the start of the object is marked, as it is sufficient for MMTk implementation, and not both the start and the end of an object. The number of pages of metadata required can be retrieved using *getRequiredMetaDataPages()* in *ChunkUtil*.

### 5. CompressorCollector

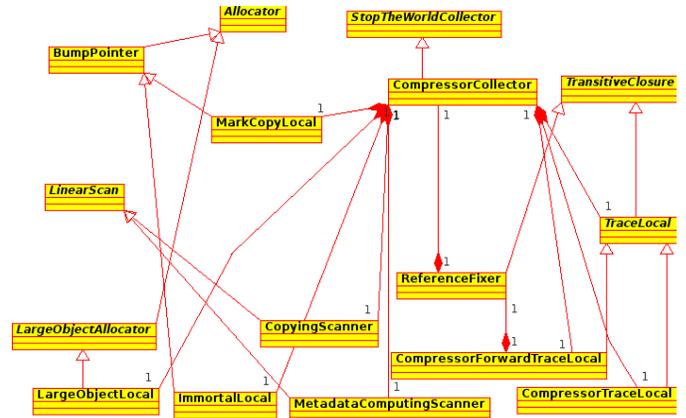

Fig. 5 Per-Collector Thread Behaviour

*CompressorCollector,* the sub class of *CollectorContext,* is the per-collector thread of Compressor, providing the GC-thread specific thread-local data structures. Fig. 5 is the class diagram of the classes interacting with CompressorCollector, that shows the per-collector thread behaviuor.

### 6. Compressor Constants

There are many constants defined and used across the classes in MMTk. *CompressorConstants* contains the public static constants that are commonly used across the compressor implementation. For ease of calculation, logarithmic arithmetic is used in the constants, along with the binary shift operators. A few examples would be,

```
public static final int LIVE_BYTES_PER_CHUNK
= 1 << LOG_LIVE_BYTES_PER_CHUNK;
public static final int LOG_BITS_IN_INT =
LOG_BITS_IN_BYTE + LOG_BYTES_IN_INT;
```

A value of 5 for *LOG_LIVE_COVERAGE* would indicate a coverage of 32.

### 7. Computation of the Data Structures

The memory map of the compressor Meta data (chunk meta data) contains four data structures as depicted below.

```
[LiveBits map | Offset Vector | First Object
Vector | Status Vector]
```

There is an entry in the offset vector, for each object that is moved to the to-virtual-space, pointing to the new location of the first object. The Offset Vector and the FirstObject Vector are calculated by CompressorCollector.

The constants deriving the offsets are defined as below, depicting the order.

```
public static final Extent LIVE_BYTES_EXTENT
= Extent.fromIntSignExtend(
LIVE_BYTES_PER_CHUNK);
public static final Extent LIVE_BYTES_OFFSET
= Extent.zero();
public static final Extent OFFSET_VEC_OFFSET
= LIVE_BYTES_EXTENT;
public static final Extent
FIRSTOBJ_VEC_OFFSET  =
OFFSET_VEC_OFFSET.plus(
OFFSET_VEC_BYTES_EXTENT);
public static final Extent STATUS_VEC_OFFSET
=FIRSTOBJ_VEC_OFFSET.plus(FIRSTOBJ_VEC_BYTES_
EXTENT);
```

The base of the *LiveBits Map, Offset Vector, First Object Vector,* and *Status Vector* are derived in the MetadataUtil, by adding the above constants using *getMetadataBase(address).plus(..)* respectively. Given an address, the address of the cell that contains the respective LiveBitsMap, OffsetVector, First Object Vector, and Status Vector can be found using, *getLiveWordAddress(), getOffsetVectorCellAddress(), getFirstObjCellAddress(),* and *getStatusVectorCellAddress()* respectively. Here the respective metadataOffset is found and added to the respective base of the metadata, to find the cell address.

The Metadata Computation extensively uses the operations available in *org.vmmagic.unboxed.Word*. The methods *liveBitSet(..)* test the live bit of the given address or the object, and returns *true* if the live bit is set. The liveness bit of a given address is set or cleared by *updateLiveBit()*. This method is called by *testAndSetLiveBits()* to atomically set the live bits for a given object or address. Similarly, *clearLiveBit()* clears it.

Here, given an ObjectReference ref, the respective address is found by

```
Address address =
VM.objectModel.objectStartRef(ref);
```

**8. BlockUtil**

*BlockUtil* defines the operations over block-granularity meta-data. Given an block address and the limit, *liveData()* calculates the live data in the block up to the address limit provided. If no limit provided, *Address.zero()* is assumed as the default and the total block size would be found. Private method *scanBlock()* is used here, to scan the block till the block end is reached. We realize that we have scanned all the objects in the block, when the current object end is equal to the block end. Here, *getBlockEnd()* is used to get the end of the block, given the start and limit addresses.

**9. Meta-data Computation**

*MetadataComputingScanner* class is a linear scanner, that takes care of all of the auxiliary meta-data computation (offset vector and first-object vector), after the marking phase. *scan()* of MetadataComputingScanner uses setOffset() and setFirstObject to set the respective offset and first object in the respective vectors.

The scanner traverses BumpPointer (BP) regions, where the compressor thread chooses a bump pointer initial region and traverses the linked list to calculate the metadata.

Since traversing the bump pointer region is done per thread by the compressor thread, interleaving of regions in the chunk in the memory is possible.

| BP1 → [1] → [2] → || BP2 → [1] → [2] → || |

After the computation, chunk in memory,

| | BP1.1 | BP 2.1 | BP 2.2 | BP 1.2 | |

As illustrated above, the address of the initial region of the Bump Pointer is pushed to the linked list appropriately. The Linked list heads are popped from the queue by the collector, PREALLOC_COPY.

The total live data of the each list is calculated. Space for the live data in the current bump pointer, bound to the to-virtual-space, is preallocated for each list. A copying scanner with the region that the object to copied to (designation), is initialised.

**10. RegionCursor**

RegionCursor is a cursor class that maintains pointer to the 3 addresses - *region,* the current region, or zero if the cursor is invalid, such as after advancing past the end of the current work list; *cursor,* the current address; and *limit,* the end of the current region. This is the address of the last used byte, when reading a populated region, and this is the last byte in the region, when writing to a fresh region. The method *isValid()* returns *true*, if we haven't passed beyond the end of the region list.

*advanceToNextRegion()* proceeds to the next region from the linked list of the space. *isAvailable()* checks whether the provided size of bytes are available in the current region.

**11. CopyingScanner**

With the *CopyingScanner*, a linear scan of the popped BumpPointer head is started using *scan()*, where the *CopyingScanner* holds a *ReferenceFixer* object and a *RegionCursor* cursor to the destination region. This advances using *advanceToNextRegion()*, till it finds the place using *isAvailable()*, to copy. Finally it copies the object.

**12. ReferenceFixer**

The scan fixes references using the precomputed metadata and copy each of the objects. *ReferenceFixer* rewrites the slot with the new to-space address, by processing every edge in an object, using the computed metadata.

*getNewReference()* gets a new reference using Compressor Metadata, for the object pointed by the provided objectReference.

**13. TraceLocal**

Classes that extend *TraceLocal* provide thread-local data structures which associates a specific way of traversing the heap. Compressor implementation has two classes extending this. *CompressorTraceLocal* implements the thread-local functionality for a transitive closure over a Compressor collector. Trace is used only to compute auxiliary metadata. The metadata is local to the a bump pointer. The order is currently maintained only in a thread local granularity. *CompressorForwardTraceLocal* implements the thread-local functionality for a transitive closure over a mark-compact space during the forwarding phase. This is used to forward only precopied objects and to fix references in these objects.

The core method of *TraceLocal* is *traceObject(ObjectReference object),* during the tracing of the object graph. This method ensures that the traced object is not collected. It also makes sure to enque the objects to be scanned, in its first visit, and returns the forwarded reference to the object. In these classes, *isLive()* checks whether the object is live.

**14. Moving the Objects**

The method *willNotMoveInCurrentCollection()* checks whether the given object will not move in the current collection. In In *CompressorForwardTraceLocal,* this check is simply checking whether they are not from the two virtual spaces, as they always move. In *CompressorTraceLocal,* as in the mark phase, the space except the from-virtual-space will not move. Objects that must not be moved can be ensured that they do not move for the rest of the GC, by using the method *precopyObject().*

### 15. Compressor Constraints

GC implementations are expected to have a *XXXConstraints* class extending the abstract classes *StopTheWorldConstraints* or *ConcurrentConstraints,* which extend the class *SimpleConstraints*, which is a sub class of *PlanConstraints,* to provide the configuration information that might be required by the host virtual machine. In order to prevent the circular dependencies, this class is separated from the Plan class.

PlanConstraints have the default values for the constraints, and the *CompressorConstraints,* as any other GC implementation, overrides them appropriately. Methods *needsForwardAfterLiveness(), needsLinearScan(),* and *movesObjects()* are overridden to return *true,* which by default return *false* in the base implementation.

### 16. Phases

Garbage collectors in MMTk work in phases, where the context of execution of the phases are global, mutator, collector, and concurrent. The method *processPhaseStack()* in the class *Phase* is called by multiple threads to process the phase stack. Each phase is either simple (*SimplePhase*) or complex, where it is a sequence of phases (*ComplexPhase*). *ConcurrentPhase* runs concurrently with the mutator activity.

The execution of the compressor algorithm is done following the same protocol, as with the other GC algorithms implemented in MMTk. For example, in a mark and sweep collector, there is no explicit calls invoking the mark and sweep phases. Rather, marking is done via tracing the heap, where tracing starts from the roots, and finishes with a transitive closure. Similarly, sweeping is done by walking through the heap and the objects that are not marked are swept.

### 17. Precopying Phase

During the precopying phase, objects are precopied to the immortal space. They are marked as forwarded and left behind a forwarding address. Then the from-virtual-space is scanned for references to precopied objects. The references with the forwarding addresses found in objects are updated. Thenceforth, all the references to precopied objects resembles the references to objects outside the MarkCopy space.

### 18. Customized Phase Schedule

Since forwarding is needed for the precopied objects, in the phase that is executed to perform a collection, compressor uses a modified phase schedule. The per-collector *collectionPhase* is implemented in *CompressorCollector*, which implements the per-collector thread behaviour and state of the Compressor. The below code segment creates a complex phase with all the phases in the customized phase schedule of Compressor.

```
public short compressorCollection =
Phase.createComplex("collection", null,

//Starts collection. Prepare collected spaces
Phase.scheduleComplex(initPhase),

/*Determines the liveness from the roots.*/
Phase.scheduleComplex(rootClosurePhase),

/*The complete closure that includes
reference types and finalizable objects.*/
Phase.scheduleComplex(refTypeClosurePhase),

/*All Mutator initial region addresses in
this phase are pushed to a global list.*/
Phase.scheduleMutator(PUSH_MUTATOR_HEADS),

//Switch to forwarding trace and prepare it
Phase.scheduleGlobal(PREPARE_FORWARD),
Phase.scheduleCollector(PREPARE_FORWARD),

/*Prepare MarkCopyLocal for the collection.*/
Phase.scheduleMutator(PREPARE),

/*SimplePhase - stacks*/
Phase.scheduleCollector(STACK_ROOTS),

/*SimplePhase - root*/
Phase.scheduleCollector(ROOTS),
Phase.scheduleGlobal(ROOTS),

/*Ensures all the references are correct.*/
Phase.scheduleComplex(forwardPhase),

/*Traverse all regions pushed in
PUSH_MUTATOR_HEADS and calculate metadata.*/
Phase.scheduleCollector(CALC_METADATA),

/*Preallocate amount needed to copy data from
each bump pointer. Copy using two cursors. */
Phase.scheduleCollector(PREALLOC_COPY),

/* Closure of the forwarding stage*/
Phase.scheduleCollector(FORWARD_CLOSURE),

/*CompleteClosure including reference types
and finalizable objects.*/
Phase.scheduleComplex(completeClosurePhase),

/*Release the forwarding trace.*/
Phase.scheduleCollector(RELEASE_FORWARD),
Phase.scheduleGlobal(RELEASE_FORWARD),

/*Completing scheduleCollector,
scheduleGlobal,and POST_SANITY_PLACEHOLDER.*/
Phase.scheduleComplex(finishPhase));
```

The phases, *initPhase, rootClosurePhase, refTypeClosurePhase, forwardPhase, CompleteClosurePhase,* and *finishPhase*, are defined as protected variables of type short in *Simple*.

### 19. MarkCopySpace and MarkCopyLocal

*MarkCopySpace* is the space that supports marking in a side mark bit similar to that in MarkSweepSpace and Copying. *MarkCopyLocal* is the thread local instance of MarkCopySpace. Fig. 6 shows how the MarkCopy(Local|

Space) objects are aggregated into the Compressor and CompressorMutator, as we discussed previously.

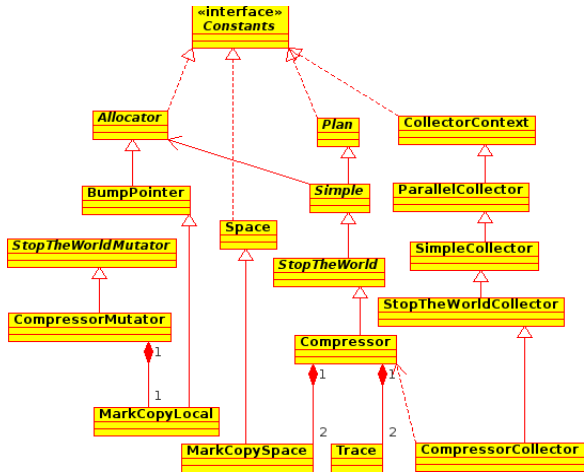

Fig. 6 Class Diagram of Compressor

*CompressorMutator* allocates objects using MarkCopyLocal with the *alloc()* and using *setInitialRegion()* to set the initial region of the cursor. But, after a collection, regions that were copied to, are used or extended to allocate new objects.

## 20. Collection Phase

*markTrace* and *forwardTrace* are two traces used during the collection phase. markTrace marks the live objects in a side markbit and precopying. The site markbit is reserved per chunk. MetadataUtil is used to access and query the markbit vector.

forwardTrace fixes all the references to the precopied objects, by scanning the from space for the references and forwarding them. While scanning in the closure phase, it also fixes references from these objects to the from-virtual-space.

## 21. Global Collection Phase

The global collection phase is executed by the *collectionPhase()* in Compressor. In the *PREPARE* phase, the roles of the virtual spaces are swapped, and the virtual spaces and the markTrace are prepared. In the *CLOSURE* phase, markTrace is prepared. In the *RELEASE* phase, markTrace and the from-virtual-space are released. Here, *fromSpace()* identifies, which space among the two is the effective from-virtual-space during this iteration. *release()* in *MarkCopySpace* releases the respective virtual space after a collection, by releasing all the pages that were associated.

In the *COMPLETE* phase, the boolean flag *readyForAllocation,* which is false by default, is set to true, for *alloc()* in *CompressorMutator*. In the *PREPARE_FORWARD* phase, the *forwardTrace* is prepared, In the *RELEASE_FORWARD* phase, forwardTrace is released, by calling *release()* in *Trace*. Here *release()* releases the resources, as the collection pass is completed.

## 22. Per-Collector Collection Phase

*CompressorCollector* contains a *TraceLocal* object *currentTrace,* to point to *markTrace*, the instance of *CompressorTraceLocal* and *forwardTrace,* the instance of *CompressorForwardTraceLocal,* alternating. It also contains a *LargeObjectLocal* instance, *largeObjectLocal*. The *allocCopy()* implementation allocates space for copying after .

CompressorCollector performs the per-collector collection using *collectionPhase()* in *CompressorCollector.* In the *PREPARE* phase, currentTrace is set to markTrace, the bump pointer is rebound to the appropriate virtual space, and the markTrace and largeObjectLocal are prepared. In the *CLOSURE* phase, *completeTrace()* of *TraceLocal* is called, which finishes processing all the GC work, iterating till all the work queues are empty. In the *CALC_METADATA* phase, the offset vector and the first object vector are calculated.

*PREALLOC_COPY* phase dequeues from the bump pointer start address space. markTrace and largeObjectLocal are released. Calculate the total live data in the last block using *MarkCopyLocal.getTotalLiveData()* and the memory is preallocated to the destination space using *MarkCopyLocal.prealloc()*. Finally the region is scanned using the *MarkCopyLocal.linearScan(),* that starts scanning from the region start, and copy to preallocated. New bumpPointer head is copied into global queue to be used by the mutators.

In the *RELEASE* phase, largeObjectLocal and markTrace are released. In the *PREPARE_FORWARD* phase, currentTrace is set to forwardTrace, and forwardTrace is prepared. Finally, in the *RELEASE_FORWARD* phase, currentTrace is set back to markTrace and the forwardTrace is released.

## 23. Per-Mutator Collection Phase

The collector threads run in parallel and iterate over the available MutatorContext objects, to run the mutator's *collectionPhase().* One of the collector threads will execute the *collectionPhase()* in *CompressorMutator,* on behalf of it. Here the collector threads perform per-mutator, where the mutator threads are mostly stopped. In the *PREPARE* phase, the MarkCopyLocal instance (markCopyLocal) is prepared, by calling *prepare()* from *MarkCopyLocal,* marking the last region in the allocation bump pointer, with its end. In the *PUSH_MUTATOR_HEADS* phase, the initial addresses of the bump pointer is pushed. Finally, in the *RELEASE* phase, the bump pointer is rebound to the appropriate virtual space.

### C. CONCURRENT IMPLEMENTATION

Most of the GC implementations of MMTk start by extending the StopTheWorld implementation. So does the Compressor algorithm. When we decided to implement the Concurrent Compressor, we decided to extend the ConcurrentCompressor. This design pattern is same as the design and development of MS (Mark and Sweep) and CMS (Concurrent Mark and Sweep) collectors.

Meta data (offset vector and first-object vector) computation can be done when the program threads are still running. With minor modifications, moving pages and updating pointers too is possible, while the program threads are running. To fix the roots, the mutator threads however should be stopped. For this, the offset vector should be computed before the program threads are stopped.

Fig. 7 shows the class diagram of the concurrent implementation. *MarkCopySpace* is extended as

*ConcurrentMarkCopySpace*. *ConcurrentCopyingScanner* extends LinearScan, copying the objects similar to CopyingScanner. *ConcurrentReferenceFixer* is the concurrent implementation of the ReferenceFixer.

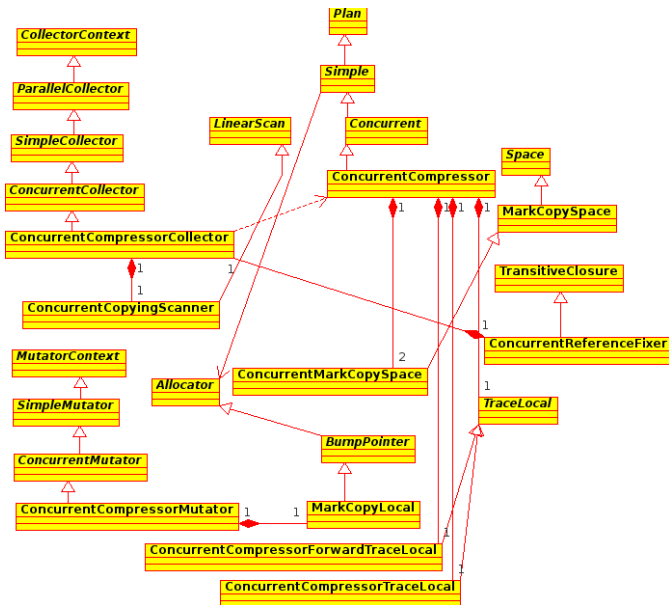

Fig. 7 Class Diagram of Concurrent Implementation

### D. DEPLOYMENT

We already had configured prototype deployments of Jikes RVM, and used it to run the Java applications, during the proposal phase, to analyse the feasibility of the project [19]. Multiple configurations for Compressor build were composed and they can be found from /build/configs/. With these configurations, the following Compressor build properties are made - *BaseBaseCompressor.properties* (Baseline compiler only build, without the optimizing compiler), *FastAdaptiveCompressor.properties* (Production), *FullAdaptiveCompressor.properties* (Development), *ExtremeAssertionsFullAdaptiveCompressor.properties* (Development build with Assertions), and *ExtremeAssertionsOptAdaptiveCompressor.properties* (Similar to the production settings, but with extreme level of assertions). The similar properties for ConcurrentCompressor too were created.

We deployed FastAdaptive compiler as well as the BaseBase Compiler, with Compressor collector and MS collector for testing. As can be seen in the *build/configs* directory, multiple versions of jikesrvm can be configured in a single machine, which is a convenience when it comes to testing. Running *bin/buildit -j $JAVA_HOME localhost FastAdaptive Compressor* from the root directory of jikesrvm will build the Fast Adaptive RVM with the Compressor Collector, in a directory such as, *dist/FastAdaptiveCompressor_x86_64-linux,* based on the architecture, following the *FastAdaptiveCompressor.properties*. Similarly Compressor can be built with the other configurations mentioned above too, by replacing *<config>* below appropriately, with *BaseBase*, *ExtremeAssertionsFullAdaptive,* or *ExtremeAssertionsOptAdaptiveCompressor*.

```
bin/buildit -j $JAVA_HOME localhost <config>
Compressor
```

Compilation is not handled by the rvm. Hence the java classes should be compiled with a JDK such as Oracle JDK. The compiled Java classes or jar archives can be run from the rvm directory, as rvm is just a virtual machine for Java.

```
./rvm -jar  xpdbench-1.0.0-jar-with-
dependencies.jar
```

### E. DEBUGGING

For the proper functionality of the new implementation of the algorithm, some other bugs or enhancements related to MMTk [17] were also addressed. These bug fixes ensured the proper integration of the new implementation, whilst letting us learn the memory management of Jikes RVM effectively.

Jikes RVM development requires debugging without proper debugger support. We may use gdb for debugging [18]. However, we decided to use the log outputs, using the *org.mmtk.utility.Log* class for debugging purposes. The below code segment is such a log statement.

```
if(VERBOSE) {
  Log.write("Using collectorHead ");
Log.write(head); Log.writeln();
}
```

Here, *VERBOSE* is defined as,

```
public static boolean VERBOSE = false;
```

and during the development and testing phase, we changed it to true in the respective classes, as required. Many of the debug statements included during the development stage were later removed, when they were no more necessary. The remaining are changed to use the verbose command line option of rvm, by including the log statements inside the below block.

```
if (VM.VERIFY_ASSERTIONS &&
Options.verbose.getValue() >= 9) { … }
```

These logs are displayed only when the RVM is run with the respective verbosity, as given below.

```
./rvm -showfullversion -X:gc:verbose=9
HelloWorld
```

Our debug statements will be printed only if the verbose level is set to 9 or more by a command line argument. Hence, for the verbosity level of 8 or less, they won't be logged. This methodology avoids logging overhead in the production builds, while enabling detailed logs for the development and debugging efforts without any code changes.

### IV. EVALUATION

Jikes RVM comes bundled with multiple plans, allowing us to build it with a preferred garbage collection algorithm, either concurrent such as concurrent mark and sweep (CMS), or the stop the world implementations that are not concurrent. When implementing a concurrent garbage collector, it is recommended to benchmark it against Mark-and-sweep collector, as it is the complete collector that has both concurrent (Concurrent Mark and Sweep - CMS) and Stop-the-World (Commonly referred to as, Mark and Sweep or MS in the code base) implementations in production with a clear separation between both implementation. The other collectors operate in the stop-the-world manner halting the program threads, where the collectors such as GenImmix, that are mostly concurrent, do not have a separate concurrent and stop-the-world implementation.

### 5.1 XPDBench

For the benchmarking purposes, a long running Java application that consumes considerable memory by creating a huge number of objects was created as a micro-benchmark. The micro-benchmark, named as *XPDBench,* was separately developed by the team, under the Apache License v2.0. XPDBench uses Apache Open Source technologies. Apache Maven is used to build and manage XPDBench by running *mvn clean install*. The subversion source code repository is hosted in SourceForge [20] along with a snapshot version of the Jikes RVM trunk, modified with our implementation. Apache Log4j is used for logging the messages to the terminal, and Apache Axiom is used to parse *xpdbench.xml*, the file that provides the configuration parameters to XPDBench.

XPDBench creates a list of double arrays with the size of elements provided, in a class named, *BigObject*, whose instances are indeed 'big objects'. For example, providing

```
<objectSize>1020</objectSize>
```

will initialize

```
bigArray = new double[1020][1020]
```

where the list contains 1020 such double array objects, as below.

```
for (int i = 0; i < 1020; i++)
{ bigList.add(bigArray); }
```

So the effect of increasing the size (n) of the object is n^3. XPDBench can be configured such that to induce garbage collection using *System.gc(),* by specifying the parameter, or letting it completely to the GC algorithm to handle it. Configurations such as how many iterations should the benchmark run, how many threads to be used running the benchmark, and how long should the tasks wait between spawning, can also be specified via the configuration file. The objects created by the application are eventually left to be garbage collected. XPDBench was initially tested for correctness against Java(TM) SE Runtime Environment (build 1.6.0_31-b04), and once finalized, was used to benchmark JikesRVM Collector implementations.

### 5.2 Environment

The efficiency of the garbage collection algorithms was measured by visualizing how effectively the memory is released and managed, for the implementations of the collectors, using XPDBench, as well as the available benchmarks. The development and evaluation was carried on a platform of Ubuntu 12.04 LTS (precise) – 64 bit, Kernel Linux 3.2.0-40-generic and GNOME 3.4.2. The environment had 1.9 GiB memory and Intel® Core™2 Duo CPU T6600 @ 2.20GHz × 2 processor available.

The default initial heap size is 50 MB and maximum is 100 MB, which can be changed by command line arguments, for the run. Given below is the configuration started with 500 MB initial, and 1000 MB maximum of heap size.

```
./rvm   -showfullversion   -Xms500M   -Xmx1000M
-jar xpdbench.jar
```

### 5.3 Benchmarks

Jikes RVM comes bundled with multiple test cases for benchmark suites and test benches, which can be found at testing/tests. For example, SPECjbb2000, SPECjbb2005, SPECjvm98, SPECjvm2008, and mmtk-harness. DaCapo [21] is an open source benchmarking suite. It is specifically used for the client side testing of the RVM. SPECjvm2008 [22] is freely available to download and use for non-commercial educational work.

For the evaluation and benchmarking, multiple initial and maximum heap sizes were used. We did various testing of the existing compressor algorithms using DaCapo and XPDBench, using the default configurations. The outcomes can be found at [23]. To run the DaCapo, SPECjvm2008, and SPECjbb2005 tests against a production build or a BaseBase Compressor build,

```
bin/buildit   localhost   -j   $JAVA_HOME   -t
dacapo   -t   SPECjvm2008   -t   SPECjbb2005
production
```

or

```
bin/buildit localhost -j $JAVA_HOME -t dacapo
BaseBase Compressor
```

### 5.4 Concurrency

When testing for concurrency, we found that Stop-the-World implementations were outperforming concurrent implementation, against the expectation, in both default heap size and bigger heap. This is due to the fact that Concurrent collectors perform collection incrementally, along with the mutator threads focusing to reduce the time the program threads are halted. This happens even where is enough free space available, and the stop-the-world implementation doesn't do collection where the mutator threads will keep running till the memory is exhausted. This provides a favourable result for the stop-the-world implementation, when the memory is available abundantly.

In case of default or minimum heap sizes, the Concurrent MS failed the eclipse benchmark of DaCapo, where it was successful with the Stop-the-World implementation, which again may be due to the implementation issues in concurrent collectors of MMTk. Due to the bugs in our Concurrent Compressor implementation, we couldn't benchmark it to confirm the issues.

### 5.5 Measurements

The FastAdaptive Compressor was benchmarked with DaCapo and XPDBench, against the existing mark-and-sweep

(ConcMS, MS) and MarkCompact (MC) FastAdaptive builds. Fig.8 depicts the time taken with -Xms500M -Xmx1000M, where 1000,10(D) depicts the values taken with the default heap size of Jikes RVM. 1000, 10 (the default settings) depicts the object size of 1000 with a thread count of 10. T (*true*) indicates whether the GC is induced by the configuration.

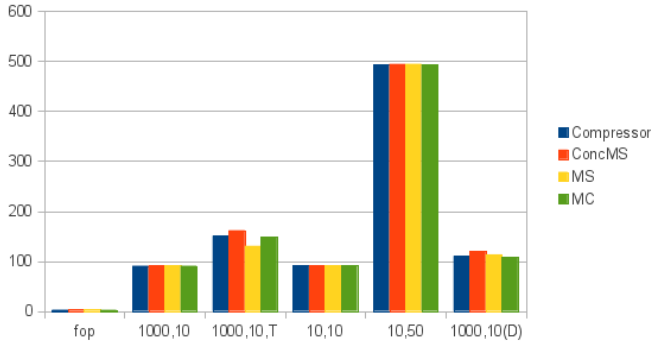

Fig. 8 Time taken in seconds.

## V. FUTURE WORK

This project focused on learning the JIT compiler and garbage collection technologies employed in Jikes RVM. The concurrent implementation is currently highly buggy and henc incomplete. It could be fixed by going through the errors and races that are caused. The original authors of the paper have proposed further optimizations to the basic algorithm to reduce the overheads in the basic algorithm, such as moving multiple pages in each trap and double-mapping to the to-virtual-space in the beginning, reducing the trap time.

Currently the implementation doesn't follow the checkstyle rules of Jikes RVM. We hope, once these short comings are fixed, the implementation could be merged to the Mercurial source code repository of Jikes RVM, which will provide an efficient concurrent and stop the world GC implementations to the Jikes RVM. These fixes and optimizations can be considered the future work of this implementation.

## VI. CONCLUSIONS

Jikes RVM has a huge code base. However, by carefully picking the modules to study and develop, the team focuses to learn some of the core concepts of virtualization technologies, while avoiding the need to learn or code for the entire code base of the project. Ideally, JIT Compiler can be independent from the VM. MMTk is very flexible that it can support multiple garbage collection algorithms, and it is pretty straight forward to introduce new garbage collection algorithms. JIT Support for memory management is an interesting topic of research, making the topics that we have picked for the analysis somewhat related.

The Compressor Mark-Compact implementation is expected to mitigate the fragmentation problem faced by the widely used Mark-and-sweep garbage collector. Other compactors require multiple heap-passes, where Compressor requires just a single pass. Copying collectors, on the other hand, do not preserve the object order, unlike Compressor. These desired properties make Compressor an effective GC algorithm.

## VII. APPENDIX: ADDITIONAL CLASS DIAGRAMS

In this section, we will go through a few additional class diagrams that our team came up with, during the internal mechanisms study of the optimization compiler.

### A. CLASS DIAGRAMS FROM GARBAGE COLLECTION

**Counters**

The class diagram depicting the counters in the package *org.mmtk.utility.statistics* is depicted in Fig. 9.

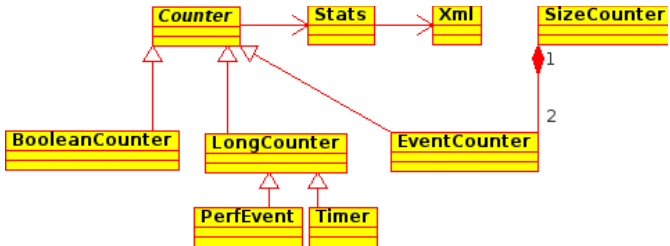

Fig. 9 Class Diagram of Counters

**Classes extending Deque**

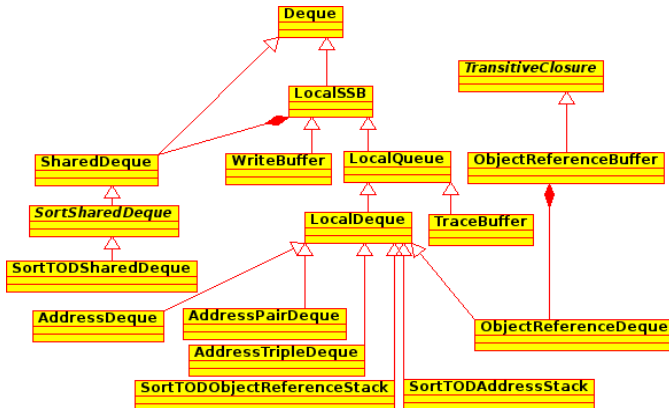

Fig. 10 Class Diagram of Deque

Fig. 10 shows the classes in the package *org.mmtk.utility.deque*. These classes extend the base class Deque.

### B. CLASS DIAGRAMS FROM OPTIMIZATION COMPILER

**Class Hierarchy of Operands**

Fig. 11 shows the class diagram of operands, the sub classes of the abstract class Operand.

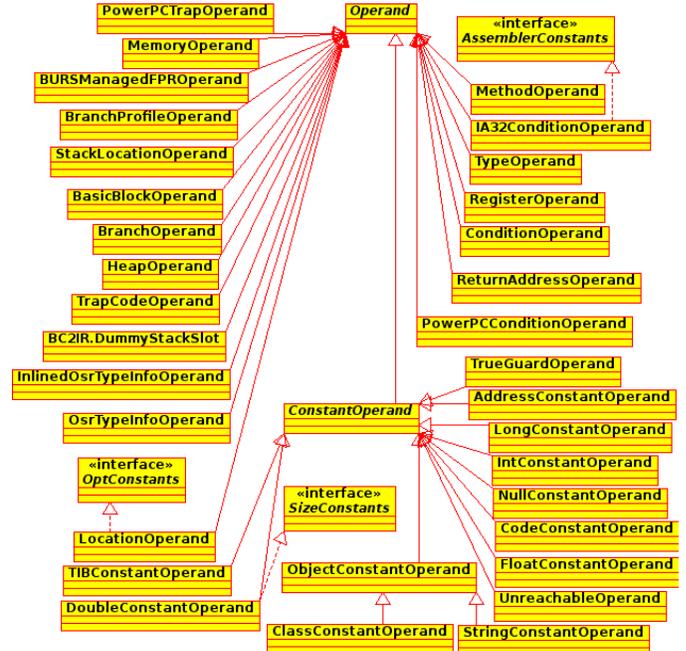

Fig. 11 Class Hierarchy of Operands

## Subclasses of CompilerPhases

Fig. 12 shows the sub classes of CompilerPhases, each responsible for different optimizations of different phases.

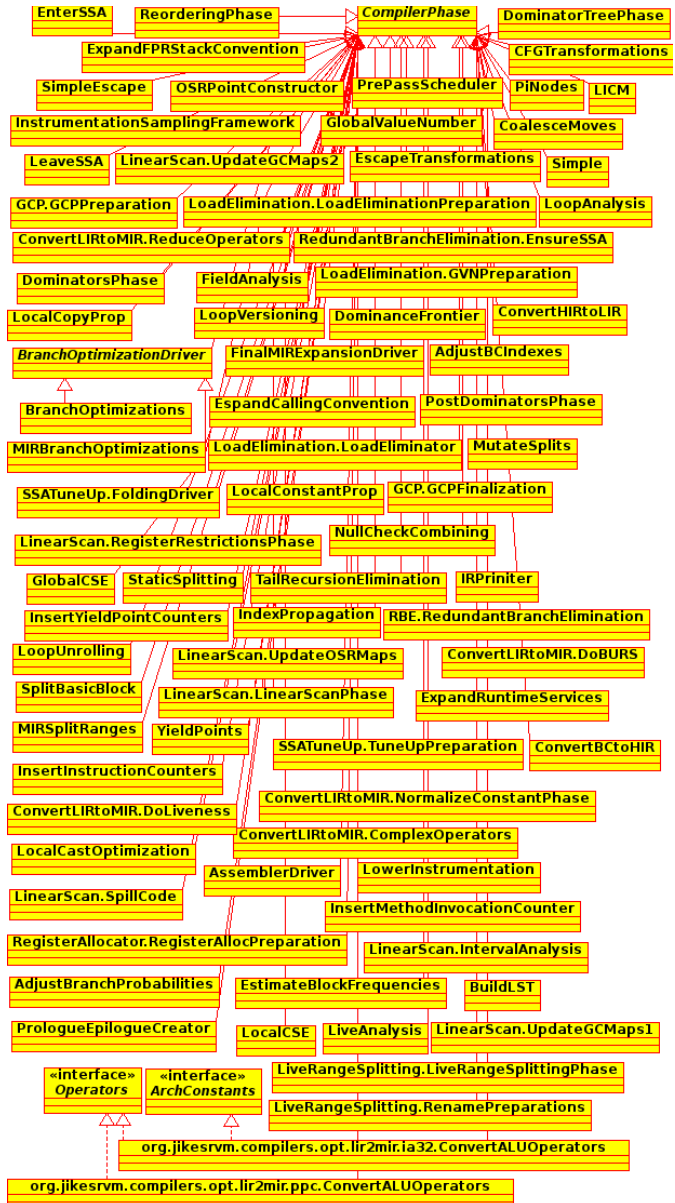

Fig. 12 Class Diagram of Compiler Phases